# MICRO-OPTICAL RESONATORS FOR MICROLASERS AND INTEGRATED OPTOELECTRONICS
*Recent advances and future challenges*

Trevor M. Benson[1], Svetlana V. Boriskina[1], Phillip Sewell[1], Ana Vukovic[1], Stephen C. Greedy[1] and Alexander I. Nosich[1,2]
[1]*George Green Institute for Electromagnetics Research, University of Nottingham, Nottingham NG7 2RD, UK;* [2]*Institute of Radio Physics and Electronics NASU, Kharkov 61085, Ukraine*

**Abstract:** Optical microcavities trap light in compact volumes by the mechanisms of almost total internal reflection or distributed Bragg reflection, enable light amplification, and select out specific (resonant) frequencies of light that can be emitted or coupled into optical guides, and lower the thresholds of lasing. Such resonators have radii from 1 to 100 μm and can be fabricated in a wide range of materials. Devices based on optical resonators are essential for cavity-quantum-electro-dynamic experiments, frequency stabilization, optical filtering and switching, light generation, biosensing, and nonlinear optics.

## 1. INTRODUCTION

Optoelectronic devices based on optical microresonators that strongly confine photons and electrons form a basis for next-generation compact-size, low-power and high-speed photonic circuits. By tailoring the resonator shape, size or material composition, the microresonator can be tuned to support a spectrum of optical (i.e., electromagnetic) modes with required polarization, frequency and emission patterns. This offers the potential for developing new types of photonic devices such as light emitting diodes, low-threshold microlasers, ultra-small optical filters and switches for wavelength-division-multiplexed (WDM) networks, colour displays, etc. Furthermore, novel designs of microresonators open up very challenging fundamental-science applications beyond optoelectronic device technologies. The interaction of active or reactive material with the modal fields of optical microresonators provides key physical models for basic research such as cavity quantum electrodynamics (QED) experiments, spontaneous emission control, nonlinear optics, bio chemical sensing and quantum information processing (Yokoyama, 1992, 1995; Yamamoto, 1993; Chang, 1996; Vahala, 2003). We shall briefly review the state-of-the-art in microresonator design tools, fabrication technologies and observed optical phenomena, and outline the challenges for future research.

## 2.  MECHANISMS OF LIGHT CONFINEMENT AND RESONATOR BASIC FEATURES

Optical resonators can be fabricated by exploiting either (almost) total internal reflection (ATIR) of light at the interface between a dielectric (e.g. semiconductor) material and the surrounding air or distributed Bragg reflection (DBR) from periodical structures such as multilayered structures or arrays of holes. The spectra of optical modes supported by microresonators are shape and size dependent. The large variety of resonator geometries that can be realized by using either or combining both of the light confinement mechanisms opens up a wide field of research and device opportunities.

For various applications it is often critical to realize a microresonator with compact size (small modal volume, V), high mode quality factor, Q, and large free spectral range (FSR). Ultra-compact microresonators enable large-scale integration and single-mode operation for a broad range of wavelengths. Q-factor is a measure of the resonator capacity to circulate and store light, and is usually defined as the ratio of the energy stored to the energy dissipated in the microresonator. In practical device applications, the high Q of the microresonator mode translates into a narrow resonance linewidth, long decay time, and high optical intensity. Wide FSR (the spacing between neighbouring high-Q resonances) is often required to accommodate many WDM channels within the erbium amplifier communications window. The ratio Q/V determines the strength of various light-matter interactions in the microresonator, e.g., enhancement of the spontaneous emission rate, and should be maximised for microlaser applications and QED experiments. However, high-Q microresonators of optical-wavelength size are difficult to fabricate, as the WG-mode Q-factors decreases exponentially with the cavity size, and thus in general the demands for a high Q-factor and compactness (large FSR, small V) are contradictory.

*Table 1.* Types and characteristics of optical microresonators.

| Resonator type | Light confinement / Dominant modes / Features | Resonator type | Light confinement / Dominant modes / Features |
|---|---|---|---|
| Microsphere 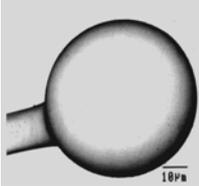 Lefevre-Seguin, V., 1999, *Opt. Mater.* **11**(2-3):153-165. © 1999 Elsevier B.V. | ATIR<br><br>WGMs<br><br>Ultra-high Q-factors ($10^7$ - $9 \times 10^9$); large mode volumes; dense modal spectrum (all modes are degenerate); challenging on-chip integration | Microtorus 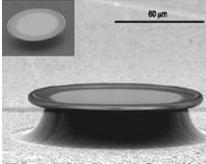 Armani, D.K. et al, 2003, *Nature* **421**:905-908. | ATIR<br><br>WGMs<br><br>Mode volumes lower than for spheres; very high Q-factors ($5 \times 10^8$); reduced azimuthal-mode spectrum; suitable for on-chip integration |
| Microdisk (microring) 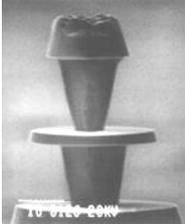 Baba, T. et al, 1997, *IEEE Photon. Technol. Lett.* **9**(7):878-880. © 1997 IEEE | ATIR<br><br>WGMs<br><br>Small mode volumes; high Q-factors ($10^4$ – $10^5$); higher-radial-order WG modes are eliminated in the ring resonators; suitable for planar integration | Quadrupolar (racetrack) microresonator 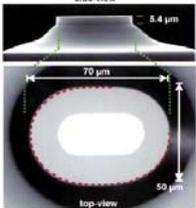 Gmachl, C. et al, 1998, *Science* **280**:1556-1564. © 1998 AAAS | ATIR<br><br>WGMs, bow-tie<br><br>Relatively low Q-factors (850-1500), highly directional emission and high FSR of the bow-tie modes; WGM Q-factors lower than in circular microdisks, efficient coupling to planar waveguides |
| Micropost/pillar 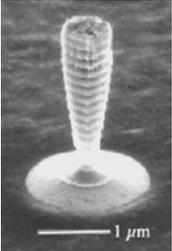 Solomon, G.S. et al, 2001, *Phys. Rev. Lett.* **86**(17):3903-3906. © 2001 American Physical Society | DBR in vertical direction<br><br>ATIR in horizontal direction<br><br>Fabry-Perot oscillations<br><br>Small mode volumes; relatively high Q-factors (1300-2000); easy coupling to fibres | Photonic crystal defect microcavity 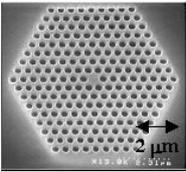 Painter, O.J. et al, 1999, *J. Lightwave Technol.* **17**(11):2082-2088. © 1999 IEEE | ATIR in vertical direction<br><br>DBR in horizontal direction<br><br>Symmetry-dependent spectrum of defect modes<br><br>The smallest mode volumes; high Q-factors ($4.5 \times 10^4$); suitable for planar integration |

A very wide range of microresonator shapes has been explored over the years for various applications (Table 1 lists some of the most popular optical microresonator types as well as their dominant modes and basic features). The most widely used are rotationally symmetric structures such as spheres, cylinders, toroids, and disks, which have been shown to support very high-Q whispering-gallery (WG) modes whose modal field intensity distribution is concentrated near the dielectric-air interface (Fig. 2). Silica microspheres exhibit the highest (nearly 9 billion) Q-factors (Braginsky, 1989; Gorodetsky, 1996; Vernooy, 1998; Laine, 2001; Lefevre-Seguin, 1997), yet have a very dense spectrum of multiple-degenerate WG modes, which complicates their application for spectral analysis or laser stabilization. Recently proposed micro-toroidal resonators (Ilchenko, 2001; Armani, 2003; Vahala, 2003; Polman, 2004) not only demonstrate very high WG-mode Q-factors approaching those of microspheres but also enable reduction of WG-mode volume, increase of resonator FSR, and on-chip integration with other components.

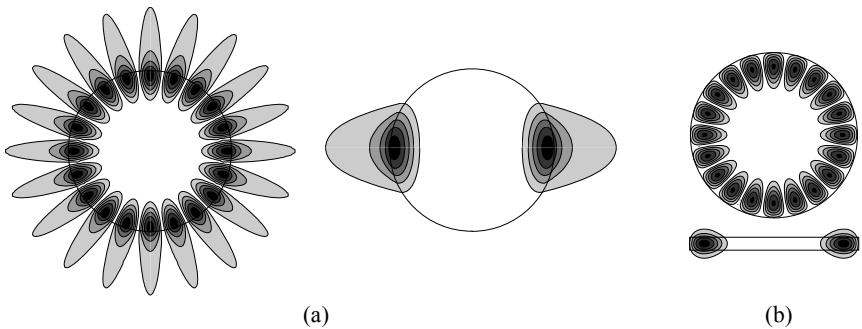

(a)  (b)

*Figure 1.* Near-field intensity portraits (16.6% contours) of the high-Q $WG_{10,1}$ modes supported by a silica microsphere (a) and a GaAs microdisk (b).

Circular high-index-contrast microring and microdisk resonators based on planar waveguide technology with diameters as small as 1-10 μm are able to support strongly-confined WG modes with typical Q-factors of $10^4$-$10^5$ and are widely used as microlaser cavities (Levi, 1993; Baba, 1997, 1999; Cao, 2000; Zhang, 1996) and add/drop filters for WDM networks (Hagness, 1997; Little, 1997, 1999; Chin, 1999). Recently, record Q-factors have been demonstrated in wedge-edge microdisk resonators (Q in excess of 1 million,

Kippenberg, 2003) and in polished crystalline microcavities ($Q>10^{10}$, Savchenkov, 2004).

Along with circular microdisk resonators, cavities of elliptical (Noeckel, 1994; Backes, 1999; Boriskina, 2003; Kim, 2004), quadrupolar (Noeckel, 1994; Gmachl, 1998; Fukushima, 2004; Gianordoli, 2000; Chin, 1999) and square (Poon, 2001; Ling, 2003; Manolatou, 1999; Hammer, 2002; Fong, 2003; Guo, 2003; Boriskina, 2004, 2005) shapes have attracted much interest. Depending on their size and degree of deformation, these microresonators can support several types of optical modes with significantly different Q-factors, near-field intensity distributions, and emission patterns (WG-like modes in square resonators (Fig. 2a), bow-tie modes in quadrupoles (Fig. 2b), distorted WG modes, volume and two-bounce oscillations, etc.). Such resonators offer advantage for various filter and laser applications as they provide splitting of the double-degenerate WG modes, directional light emission and more efficient microresonator-to-straight-waveguide coupling.

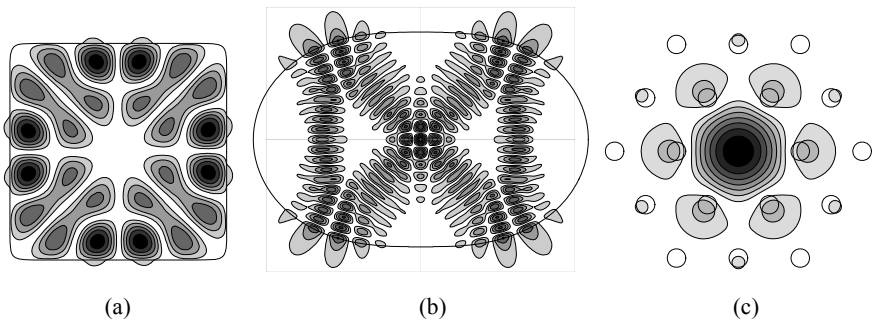

(a)    (b)    (c)

*Figure 2.* Near-field intensity portraits of a WG-like mode in (a) a square microdisk resonator and (b) a bow-tie mode in a quadrupolar (stadium) resonator and (c) a monopole mode in a hexagonal photonic crystal defect cavity (Boriskina, 2005).

Planar photonic crystal (PC) microcavities (Fig. 2c), formed, e.g., as arrays of air holes etched into a slab, have been demonstrated to simultaneously exhibit high Q-factors and ultra-small, wavelength-scale modal volumes (Foresi, 1997; Benisty, 1999; Chow, 2000; Yoshie, 2001; Kim, 2002; Akahane, 2003; Srinivasan, 2004; Noda, 2000; Painter, 1999; Boroditsky, 1999). In these cavities, the photonic-bandgap effect (fulfilment of the Bragg reflection conditions for all the propagation directions in a certain frequency range) is used for strong light confinement in the cavity plane, and TIR, for light confinement at the air-slab interface (Russel, 1996; Krauss, 1999). Modern fabrication technologies enable precise control of the PC cavity geometry, and the inherent flexibility in hole shape, size, and pattern makes fine-tuning of the defect mode wavelengths, Q-factors, and emission patterns possible (Painter, 2001; Coccioli, 1998).

In micropost and micropillar resonators (Gayral, 1998; Solomon, 2001; Pelton, 2002; Benyoucef, 2004; Santori, 2004), the transverse mode confinement is due to TIR at the semiconductor-air interface, while confinement in the vertical direction is provided by a pair of distributed Bragg reflectors. Thus, these microresonator structures can be seen as 1-D PC defect cavities in a fibre. They support Fabry-Perot-type modes with relatively high Q-factors and small modal volumes, which makes them promising candidates for microlaser applications and the observation of cavity-QED phenomena.

Other types of optical microcavities employing the DBR mechanism of light confinement include planar annular Bragg resonators (Scheuer, 2005), based on a radial defect surrounded by Bragg reflectors, and their 3-D equivalent, spherical Bragg "onion" resonators (Liang, 2004).

## 3. RESONATOR MATERIAL SYSTEMS

Progress in the design and fabrication of high-quality optical microresonators is closely related to the development of novel optical materials and technologies. The key material systems used for microresonator fabrication include silica, silica on silicon, silicon, silicon on insulator, silicon nitride and oxynitride, polymers, semiconductors such as GaAs, InP, GaInAsP, GaN, etc, and crystalline materials such as lithium niobate and calcium fluoride. Table 2 summarises the optical characteristics of these materials (see Eldada, 2000, 2001; Hillmer, 2003; Poulsen, 2003 for more detail).

Silica glass is a widely used resonator material, which combines the advantage of a very low intrinsic material loss, a large transparency window, and compatibility with standard fibre-optic technologies, e.g., efficient coupling to optical fibres (Sandoghdar, 1996; Gorodetsky, 1996; Laine, 2001; Cai, 2000; Ilchenko, 2001). The silica-on-silicon technology that involves growing silica layers on silicon substrates is one the most widely used technologies to fabricate planar microresonator devices (Eldada, 2001; Kippenberg, 2003; Armani, 2003; Polman, 2004), although microring resonators made of compound glasses ($Ta_2O_5$-$SiO_2$) have also been demonstrated (Little, 1999). A glass platform is relatively inexpensive, and as the index contrasts are smaller than for semiconductors, larger-size single-mode resonators can be fabricated by using cheaper lithographic techniques. However, wavelength-scale resonators with wide FSR in high-index-contrast silicon or silicon-on-insulator material systems (Little, 1998; Akahane, 2003; Dumon, 2004) are crucial for ultra-compact integration of photonic integrated circuits. Another relatively new planar resonator platform is based on using silicon nitride and oxynitride (Krioukov, 2002; Melloni, 2003; Barwicz, 2004). It enables the index contrast to be adjusted by as much as 30% by changing the resonator material composition in between that of $SiO_2$ and $Si_3N_4$.

*Table 2.* Properties of some key optical microresonators material systems.

| Material system | λ emission | Refractive index at 1.55μm |
|---|---|---|
| Fused silica ($SiO_2$); Silica on silicon | | 1.44-1.47 |
| Silicon on insulator (SOI) | | 3.4757 |
| Silicon oxynitride ($SiO_xN_y$) | | 1.44-1.99 |
| Polymers | | 1.3-1.7 |
| Gallium Arsenide (GaAs) | 0.8-1.0 μm | 3.3737 |
| Indium Phosphide (InP) | 1.3-1.7 μm | 3.1 |
| Gallium Nitride (GaN) | 0.3-0.6 μm | 2.31 |
| Lithium Niobate ($LiNbO_3$) | | 2.21(2.14) |

Of course, GaAs, InP, and GaN- based semiconductors have attracted much attention as the materials for microresonator device fabrication (Baba, 1997; Hagness, 1997; Solomon, 2001; Gianordoli, 2000; Kim, 2002; Painter, 1999; Grover, 2001, 2004; Zhang, 1996; Kneissl, 2004), as they enable very compact resonator structures that can perform passive (coupling, splitting and multiplexing) and active (light generation, amplification, detection, modulation) functions in the same material system. Most semiconductor microdisk and planar PC resonators consist of thin heterostructure layers. Such structures enable local modification of the energy-band structure of the semiconductor (quantum wells) and thus control of the material emission wavelength (Einspruch, 1994). Recently, there has been a lot of interest in semiconductor resonators with artificially engineered material band structure based on low-dimensional heterostructures, such as one-dimensional quantum wires and zero-dimensional quantum boxes/dots (Arakawa, 1986).

Polymer materials (Eldada, 2000; Rabiei, 2002, 2003; Chao, 2003, 2004) offer great potential for use in advanced optoelectronic systems as they offer low material costs, up to 35% tunability of the refractive index contrast, excellent mechanical properties, very low optical loss, large negative thermo-optic coefficient and low-cost high-speed processing. Other attractive materials for high-Q microresonators fabrication, which combine a wide optical transparency window, a good electro-optic coefficient and nonlinearity, are crystalline materials such as calcium fluoride and lithium niobate (Ilchenko, 2002; Savchenkov, 2004, 2005).

In addition to semiconductor materials discussed above, there are other material systems in which electroluminescence has already been demonstrated, even in the visible wavelength range. These materials include:

microporous silicon fabricated by electrochemical etching of crystalline silicon wafers in a hydrofluoric acid (Chan, 2001), Erbium-doped silicon (Polman, 2004; Hillmer, 2003; Gardner, 2005), phosphate glass (Cai, 2000) and organic materials, e.g., para-phenylene-vinylene (Krauss, 1999; Hillmer, 2003). Furthermore, polymers that are not intrinsically functional can be doped with organic laser dyes, rare-earth light-amplifying complexes, and electro-optic dyes. Such materials can be used to fabricate laser microresonators and have an advantage over III-V semiconductors in either compatibility with silicon microelectronics or cost and ease of fabrication.

## 4.    FABRICATION TECHNOLOGY

Recent advances in nanofabrication technology offer the possibility of manufacturing novel optical microresonator devices with dimensions of the order of the optical wavelength in a variety of natural and artificial material systems. Integrated micro-ring and micro-disk resonators as well as PC defect microcavities are usually micro-fabricated on wafer substrates using well-developed integrated-circuit deposition, lithographic and etching techniques (Elliott, 1989). During the fabrication process, wafers are typically cycled through three main steps: wafer preparation, lithography, and etching. Some of these steps may be repeated at various stages of the fabrication process. A schematic drawing of a typical micro-resonator fabrication sequence is presented in Fig. 3.

The first stage of the fabrication process (wafer preparation) usually involves the growth or deposition of material layers of different composition to create vertical layered structures able to support guided waves. The layered structure might be formed using various wafer growth techniques, such as molecular beam epitaxy, chemical or physical vapour deposition, or wafer bonding, depending on the material system. The choice of the growth technique depends on the desired resonator structure, material system, and required precision. For example, semiconductor materials usually require several successive growths to make the vertical structure, while polymer layers can simply be sequentially spun onto a substrate.

Molecular beam epitaxy (MBE) is an expensive yet widely used technique for producing epitaxial layers of metals, insulators and III-V and II-VI based semiconductors, both at the research and the industrial production level (Herman, 1996). It consists of deposition of 'molecular beams' of atoms or clusters of atoms, which are produced by heating up a solid source, onto a heated crystalline substrate in ultra-high vacuum. MBE is characterized by low growth temperatures and low growth rates and thus enables producing high-precision epitaxial structures with monolayer control. Other wafer growth techniques include chemical vapour deposition (Foord, 1997) of reactant gases broken down at high temperatures, which can be used to deposit GaN, GaInAsP, Si, $SiO_2$, $Si_3N_4$, and polymer materials; sputter (or physical vapour) deposition, which is a process of controlled deposition of energetic ionized particles of $SiO_2$, $Si_3N_4$, metals, and alloys (Little, 1999); and wafer bonding, or fusing of two materials (e.g., Si to glass, Si to Si, III-V semiconductors to SiN or $SiO_2$) at the atomic level (Tishinin, 1999).

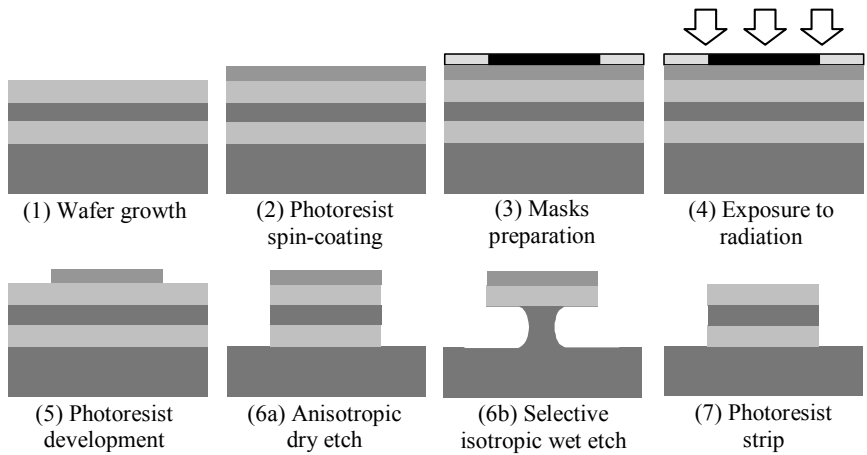

*Figure 3.* Schematic procedure of microdisk resonators fabrication.

A resonator structure is then imaged onto a wafer through a multi-step lithography process. To make the wafer (usually covered by a thin oxide film) sensitive to an image, a photoresist is spread on the wafer by a process called spin coating. Then, a photomask (a glass emulsion plate with a resonator pattern) is placed on top of the wafer. Light is projected through the voids in the photomask and images the mask pattern on the wafer. Several types of lithography with various resolutions can be used depending on the resonator type, size, and fabrication tolerances: optical ($\lambda$=160-430 nm; feature size ~ 0.13 μm), extreme ultraviolet ($\lambda$=13 nm; feature size ~ 45 nm), X-ray ($\lambda$=0.4-4 nm; feature size ~ 25 nm), or electron (ion) beam ($\lambda$=0.03 Å; feature size ~ 10-20 nm). When exposed to light, the resist either polymerizes (hardens) (if a negative resist is used) or un-polymerizes (if a positive resist is used). After exposure, the wafer is developed in a solution to dissolve the excess resist.

Once the resist has been patterned, the selected regions of material not protected by photoresist are removed by specially designed etchants, creating the resonator pattern in the wafer. Several isotropic (etching occurs in all directions at the same etch rate) and anisotropic (directional) etching processes are available. Wet chemical etching is an isotropic process that uses acid solutions to selectively dissolve the exposed layer of silicon dioxide. However, to produce vertical resonator sidewalls, anisotropic dry etching is required. In the most commonly used dry etching technique, plasma etching, ions reacting with atoms of the wafer are created by generating plasma by rf discharge. Dry etching techniques also include sputter etching, ion milling, reactive etching, and reactive ion beam etching. Another type of anisotropic etching technique is anodic etching - wet etching with an applied electric field. After the etching process, the remaining photoresist is removed. In most cases, several etching techniques are combined to form a desired resonator structure. For example, in the fabrication of microdisk resonators mounted on posts (Baba, 1997) or 2-D PC microcavities (Painter, 1999; Hennesy, 2003), the pattern is first transferred into a heterostructure by an anisotropic dry etch and then a microdisk or a PC cavity membrane is released by selective wet etching of the substrate.

Microsphere resonators can also be manufactured with standard wafer processing technologies. Usually, silicon posts topped with silica blocks are created on the wafer and then are molten into spherical shapes by controlled heating. However, the most common technique to fabricate silica microsphere resonators of several hundred microns in diameter is to simply melt a tip of an optical fibre by hydrogen flame or electric arc heating (Braginsky, 1989; Sandoghdar, 1996; Gorodetsky, 1996; Laine, 2001). To fabricate microspheres of smaller diameters, the fibre can first be thinned by tapering or etching (Cai, 2000). In the heating process, once the silica temperature passes the melting point, the surface tension forces shape the fused silica into a near-perfect spherical form. After the sphere is removed from the flame, solidification occurs almost instantaneously. The resulting microsphere has extremely smooth surface with Angstrom-scale surface deformations resulting in very high Q-factors of the WG modes.

Similar technique can also be used to fabricate silica microtorus resonators by compressing a small sphere of low-melting silica glass between cleaved fibre tips (Ilchenko, 2001). The combined action of surface tension and axial compression results in the desired toroidal geometry. Recently, a process for producing silica toroidal-shaped microresonators-on-a-chip with Q factors in excess of 100 million by using a combination of lithography, dry etching and a selective reflow process have been demonstrated (Armani, 2003). By selectively heating and reflowing a patterned and undercut microdisk with the use of a $CO_2$ laser a toroidal resonator with the atomically smooth surface was obtained.

The thermal-reflow process (or hot embossing) has also been used to reduce significantly sidewall surface roughness in polymer (polystyrene) microring resonators (Chao, 2004). This is a very useful technique because it enables post-fabrication fine-tuning of microresonator performance. However, for the fabrication of complex, multilayer devices this procedure is less useful as the high temperatures needed to reflow the polymer film could also disturb lower cladding layers and possibly alter optically active dopant molecules.

A need for submicron-scale patterning as well as stringent etching tolerances of conventional fabrication techniques often result in either very high fabrication costs or high scattering losses and frequency detuning in microresonator-based devices. An emerging lithographic technology that can achieve sub-10 nm pattern resolutions beyond the limitations set by the light diffractions or beam scatterings in the conventional methods is a nanoimprint technique (Chou, 1996; Guo, 2004). Based on the mechanical embossing principle, nanoimprint technology is used not only to create resist patterns as in lithography but also to directly pattern microresonator structures in polymers. Nanoimprinting utilizes a hard mould (which plays the same role as the photomask in photolithography) with predefined nanoscale features to mechanically imprint into a heated polymer film. The created thickness contrast pattern is `frozen' into the polymer during a cooling cycle.

Recently, as processing techniques for the synthesis of monodispersed nanoparticles of various shapes have matured, considerable effort has been directed to study the mechanisms of their self-assembly as an inexpensive and fast method of fabrication of photonic structures. Electrostatic self-assembly is an emerging technology that can create microresonator structures with tailored optical properties. For example, self-assembled arrays of polystyrene microspheres coupled to each other and to rib waveguides are expected to or already find application in wavelength selection (Tai, 2004), waveguiding (Astratov, 2004), optical sensing, and optical delays. Finally, recent advances in polishing techniques have allowed the fabrication of large (~5 mm-diam & 100-micron thick) disk and toroidal microresonators of crystalline materials, such as lithium niobate and calcium fluoride (Ilchenko, 2002; Savchenkov, 2005) that find use as high-efficiency microwave and millimetre-wave electro-optical modulators.

## 5. DEVICES AND APPLICATIONS

It is difficult to overestimate the growing importance of optical microresonators in both fundamental and applied research, and thus it is hardly possible to cover in detail the progress in all the areas of their application. In the following sections, we shall briefly review some of the major existing and emerging microresonator applications and outline the challenges calling for novel resonator geometrical designs, spectral characteristics, material properties, or robust fabrication procedures.

## 5.1    Integration with other components

One of the most difficult challenges in the design and fabrication of integrated microresonator-based photonic devices and systems is the efficient coupling of light into and out of a microresonator without compromising its narrow resonance linewidth. Furthermore, the cost and robustness of fabrication, simplicity of the microresonator-to-coupler alignment as well as ability to provide on-chip integration are very important factors in the development of advanced microresonator couplers. The most widely used microresonator coupling devices are evanescent-field couplers of various geometries such as prisms, tapered fibres, planar and PC waveguides, etc. (see Fig. 4).

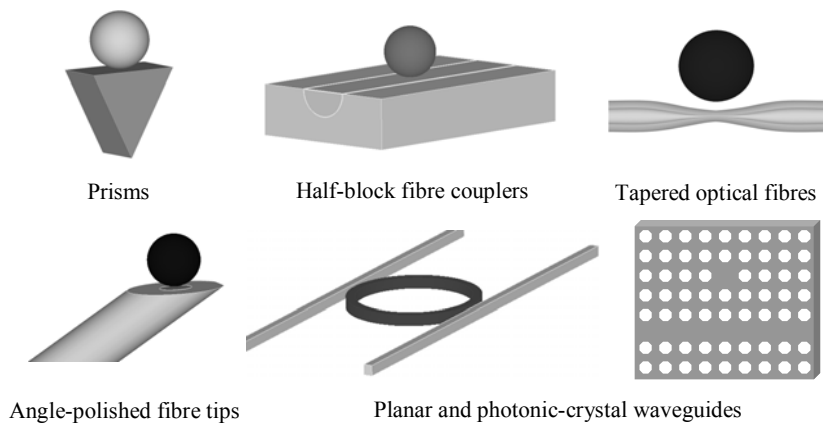

Prisms            Half-block fibre couplers        Tapered optical fibres

Angle-polished fibre tips     Planar and photonic-crystal waveguides

*Figure 4.* Various types of microresonator coupling devices.

The efficiency of optical power transfer in/out of a microresonator can be controlled by manipulating the overlap between the resonator and coupler mode fields, matching of the mode propagation constants, or changing the length of the evanescent-field coupling region. Prisms have traditionally been used to couple light into microspheres (Gorodetsky, 1996; Vernooy, 1998) and more recently, to square resonators (Pan, 2003), although they require bulk optics for focusing and alignment of the light source. Coupling to microsphere (Serpenguzel, 1995; Cai, 2000; Spillane, 2002) and microtoroid resonators (Kippenberg, 2004) through tapered fibres has several distinct advantages such as the built-in coupler alignment, relatively simple fabrication, possible on-chip integration, and control of the coupling efficiency by the change of the fibre thickness. However, integrated, wafer-fabricated microresonators are usually coupled into a photonic circuit through either planar (rib) waveguides (Hagness, 1997; Little, 1997, 1999; Chin, 1999; Grover, 2004; Zhang, 1996; Melloni, 2003) or PC waveguides (Noda, 2000).

The very small size and strong optical confinement of integrated planar microdisk and microring resonators, which make them promising candidates for large scale integration, also make them very sensitive to fabrication errors that can drastically spoil coupling efficiency. For example, if a circular microdisk is laterally coupled to a waveguide via a submicron-width air gap, the coupling efficiency strongly depends on the gap width. Accurate and repeatable fabrication of such narrow gaps by lithographic and etching techniques is a rather challenging task. The smaller the microdisk, the more difficult it is to control the coupling coefficient. For 2–5 μm-sized microdisks, the resolution of the fabrication method is often not high enough to achieve the desired narrow air gaps (<0.1 μm), and to enhance coupling, microdisks are fused to bus waveguides (Grover, 2004). Efficient lateral coupling across wider air gaps can be achieved by increasing the coupling interaction length by either curving the adjacent waveguide along the microdisk (Zhang, 1996; Hagness, 1997; Chin, 1998) or using elliptical, racetrack or square microresonators (Chin, 1999; Boriskina, 2003; Manolatou, 1999; Hammer, 2002; Fong, 2003). It has been also predicted that resonator-waveguide coupling may be enhanced by exploiting the higher field concentration at the increased-curvature portion of an elliptical microdisk resonator (Boriskina, 2003).

## 5.2 Wavelength-selective components for WDM systems

Microdisk, microring, and PC defect resonators are versatile building blocks for very large scale integrated photonic circuits, as they are ultra compact ($10^5$ devices/cm$^2$) and can perform a wide range of optical signal processing functions such as filtering, splitting and combining of light, switching of channels in the space domain, as well as multiplexing and demultiplexing of channels in the wavelength domain.

Wavelength-selective bandstop (Fig. 5a) or add/drop (Fig. 5b) filters that can combine or separate different wavelengths of light carrying different information are essential components for controlling and manipulating light in optical transmission systems. High-Q microdisks/microrings evanescently coupled to bus waveguides have been extensively explored for WDM channel dropping due to their ability to select a single channel with a very narrow linewidth (Hagness, 1997; Little, 1997, 1999; Chin, 1999; Boriskina, 1999; Grover, 2004). To compromise between the requirements for a narrow channel linewidth, structure compactness, and wide FSR, the radii of the microresonators used in filters are usually in between 5 and 30 micron, resulting in filters with a FSR of 20-30 nm. By using PC defect microcavities, further miniaturization of optical filters can be achieved (Fan, 1998). PC-microcavity filters can be realized by either introducing a single defect in the vicinity of the PC waveguide (Noda, 2000) or by integrating PC defect microcavities directly into a submicrometre-scale silicon waveguide (Foresi, 1997; Chow, 2004).

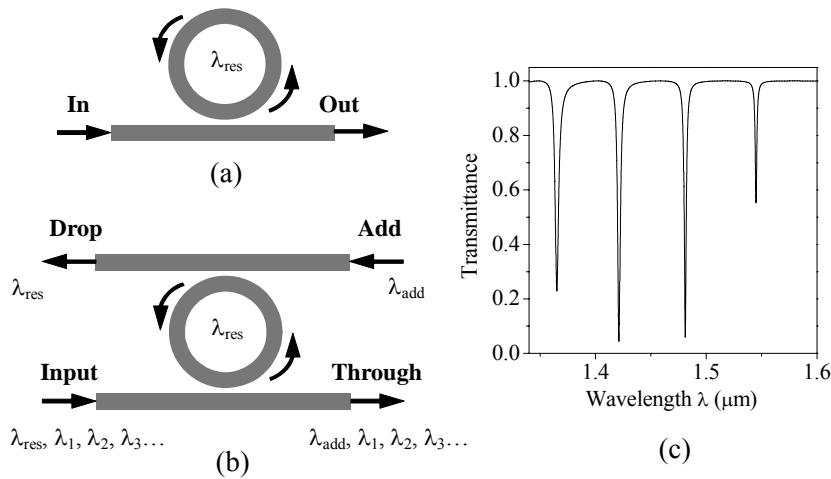

*Figure 5.* Schematics of (a) a bandstop (all-pass) and (b) add/drop filter designs, and (c) a transmission characteristic of a 3.0-μm-diameter GaAs microring resonator coupled to a bus waveguide.

An ideal filter should have a box-like spectral response with a flat passband, sharp roll-off from passband to stop band, and large out-of-band rejection. However, the transmission characteristic of a single microresonator is a series of Lorentzian-shape sharp resonance peaks at wavelengths corresponding to the excitation of the high-Q modes in the microresonator (Fig. 5c). For closely spaced channels, however, a Lorentzian response may not provide adequate roll-off to minimize cross-talk between different channels. To overcome this limitation, higher-order filters formed by cascading multiple resonators have been introduced (Little, 1997; Hryniewicz, 2000; Grover, 2002; Melloni, 2003; Savchenkov, 2003, 2005). Such filters have flat responses around resonances, much faster roll-off, and larger out-of-band signal rejection. Vernier filter tuning by combining microring resonators of different radii has also been successfully used to suppress non-synchronous resonances of different microrings and extend the resulting filter FSR (Yanagadze, 2002).

Optical characteristics of active microresonators (and thus spectral properties of the resonator-based filters) can be dynamically tuned by changing their refractive indices, for example, by using the thermo-optic or electro-optic effects. Tunable microring resonators have been used to demonstrate optical modulation using the electro-optic effect in polymer (Rabiei, 2002), all-optical switching using free-carrier injection in GaAs–AlGaAs (Ibrahim, 2003), and absorption-induced wavelength switching and routing (Little, 1998). Furthermore, significant enhancement of non-linear effects in microresonators made possible the demonstration of Kerr nonlinear phase shift (Heebner, 2004) and optical wavelength switching and conversion (Absil, 2000; Melloni, 2003) in microresonator structures.

Linear arrays of optical microresonators evanescently coupled to each other can also be used for optical power transfer (Fig. 6). This type of coupled-resonator optical waveguide (CROW) has recently been proposed (Yariv, 1999) and then demonstrated and studied in a variety of material and geometrical configurations, such as sequences of planar microrings (Poon, 2004), arrays of coupled microspheres (Astratov, 2004), and chains of photonic crystal defect cavities (Olivier, 2001).

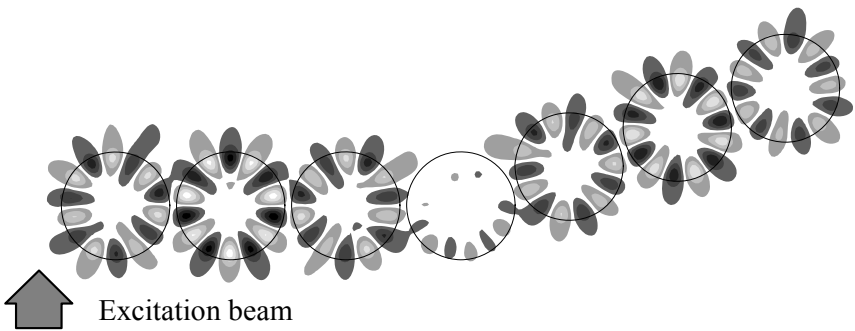

*Figure 6.* Near-field pattern inside a short chain of evanescently coupled microdisks excited by a directional beam grazing a rim of the left-hand-side resonator (Boriskina, 2005).

Among the advantages offered by CROWs is the possibility of making reflection-less waveguide bends as well as a significant slowing of light pulses. Reduction of the light group velocity in CROWs can be explored in a variety of optical applications such as delay lines, optical memory elements, and components for nonlinear optical frequency conversion, second-harmonic generation and four-wave mixing (Poon, 2004; Melloni, 2003; Heebner, 2004; Mookherjea, 2002).

## 5.3     Biochemical sensors

Microresonators supporting high-Q modes also have great potential in the development of inexpensive, ultra-compact, highly sensitive and robust bio- and chemical sensors on a chip. As compared to linear optical waveguide biosensors, microresonator-based devices benefit from much smaller size (several μm rather than a few centimetres) and higher sensitivity. Several phenomena can be used for detection. For example, cylindrical microresonators, operating on the WG modes, have been shown to enhance the intensity of the fluorescence emitted by biological materials (Blair, 2001). However, most microresonator sensors rely on the measurement of transmission or scattering characteristics of a microresonator exited by an optical waveguide mode in the presence of biological material on the resonator surface or in the surrounding solution. These sensors can detect the resonance frequency shift caused by the change of the resonator effective refractive index or increased absorption (Boyd, 2001; Krioukov, 2002; Vollmer, 2003; Chao, 2003). Alternatively, the sensors can measure the change in phase and/or intensity of the light at the output of the waveguide in the forward direction at a fixed wavelength (Rosenblit, 2004).

   High Q-factors of microresonator modes are crucial for achieving high sensitivity of the sensors. Indeed, the higher the Q-factor, the steeper the slope between zero and unity in the transmission characteristic of a microresonator, resulting in higher sensor sensitivity. Another way to enhance the slope between the zero and the unity transmission (and thus improve the sensor sensitivity) is to use a microresonator structure that generates a sharp asymmetric Fano-resonance line shape (Chao, 2003).

## 5.4     Spontaneous emission control, novel light sources and cavity QED

Another major application for microresonators is in development and fabrication of novel light sources such as resonant-cavity-enhanced light-emitting diodes (LEDs), low-threshold microlasers, and colour flat-panel displays. In wavelength-sized microresonator structures, semiconductor material luminescence can be either suppressed or enhanced, and they also enable narrowing of the spectral linewidth of the emitted light (Haroche, 1989; Yokoyama, 1992; Yamamoto, 1993; Krauss, 1999; Vahala, 2003).

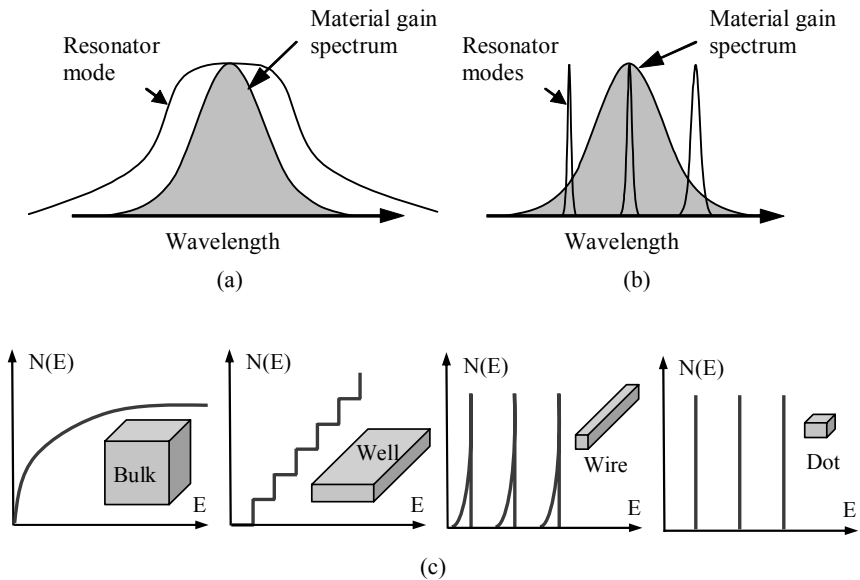

*Figure 7.* Difference in the spontaneous emission enhancement in a LED (a) and a microcavity laser (b); Density of electronic states in bulk semiconductor material and low-dimensional semiconductor heterostructures (c).

Since 1946, when it was first proposed that the spontaneous emission from an excited state of an emitter can be significantly altered if it is placed into low-loss wavelength-scale cavity (Purcell, 1946), various microresonator designs for efficient control of spontaneous emission have been explored including microdisk (Levi, 1993; Baba, 1997, 1999; Backes, 1999; Cao, 2000; Fujita, 1999, 2001, 2002; Zhang, 1996), microsphere (Cai, 2000; Shopova, 2004; Rakovich, 2003) and micropost (Pelton, 2002; Reithmaier, 2004; Santori, 2004; Solomon, 2001; Gayral, 1998) resonators as well as PC defect cavities (Painter, 1999; Boroditsky, 1999; Gayral, 1999). Depending on the Q-factors (resonance linewidths) of the modes supported by the microresonator in the spontaneous emission range of the resonator material, the two situations illustrated in Fig. 7 can be realized. In a cavity-enhanced LED (Fig. 7a), the material gain bandwidth is smaller than the cavity resonance bandwidth, and emission of the whole material gain spectrum is enhanced. In a semiconductor laser (Fig. 7b) with a laser microcavity supporting high-Q modes, the emission at one (or several) of the modal wavelengths is strongly enhanced (being predominantly stimulated emission), while the emission at all other wavelengths is suppressed. The amount by which the spontaneous emission rate is enhanced for an emitter on resonance with a cavity mode is characterized by the Purcell factor, which is proportional to the mode Q-factor and inversely proportional to the mode volume.

Clearly, to increase the enhancement factor, it is necessary to design and fabricate high-Q, small-V microresonators. However, cavity-enhanced LEDs based on the microresonators with high-Q modes must have equally narrow material spontaneous emission linewidths (Fig. 7a), which are not easily realized in bulk or heterostructure quantum-well microresonators. The recently proposed concept of an active material system, semiconductor quantum dots (QDs) (Arakawa, 2002) combines the narrow linewidth normally associated with atomic emitters and the high gain achievable in semiconductors. The QD emission spectrum exhibits delta-function-like lines with ultra-narrow linewidths (Fig. 7c). QDs of various sizes can now be fabricated by self-assembly, and have been integrated as emitters (so-called artificial atoms) in microdisk (Cao, 2000), PC defect (Yoshie, 2001; Hennesy, 2003), micropost/micropillar (Santori, 2004; Solomon, 2001; Gayral, 1998; Benyoucef, 2004), and microsphere (Shopova, 2004; Rakovich, 2003) resonators. The challenge is then to carefully design and tune the microresonator modal and geometrical properties to manipulate and increase the strength of coupling between the QD and an optical field of the resonator mode (Pelton, 2002; Hennesy, 2003).

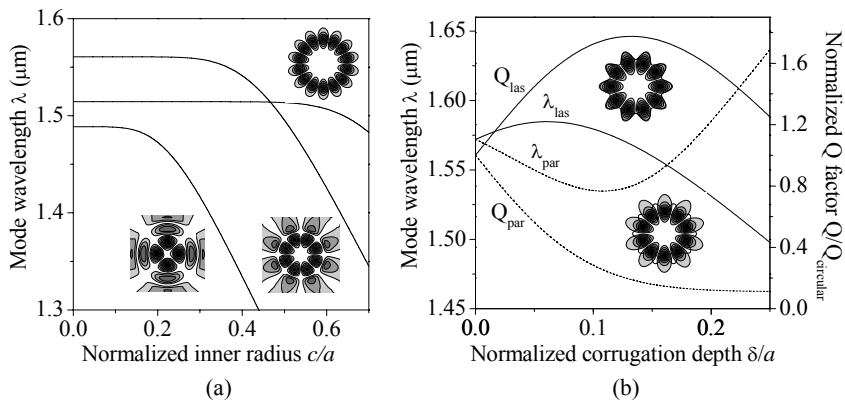

*Figure 8.* (a) Detuning of the higher-radial-order WG modes in a microring resonator of radius *a* with an increase of the inner radius *c* (Boriskina, 1999; 2002); and (b) splitting of a double-degenerate WG mode in a microgear resonator with simultaneous enhancement of the Q-factor of the working mode and suppression of the parasitic mode (Boriskina, 2004).

In the microcavity lasers, only a small portion of the spontaneous emission couples into a single (or several) optical modes (Fig. 7b). It should also be noted that optical modes of rotationally symmetrical microresonators are either multiple (microspheres) or double (microdisks, microrings, circular micropillars, etc.) degenerate. This often leads to the appearance of

*Micro-optical Resonators* 57

closely located doublets in their lasing spectra due to fabrication errors (sidewall roughness and shape imperfections, etc.) and thus causes spectral noise, mode hopping and polarization instabilities (Gayral, 1998; Fujita, 1999; Boriskina, 2004). Among recently proposed modified resonator designs that remove the mode degeneracy and help separate two closely located modes are elliptical (Gayral, 1998; Boriskina, 2003; Kim, 2004), square (Guo, 2003; Boriskina, 2005), microgear (Fujita, 2001, 2002; Boriskina, 2004), and notched (Backes, 1998; Boriskina, 2005) microresonators.

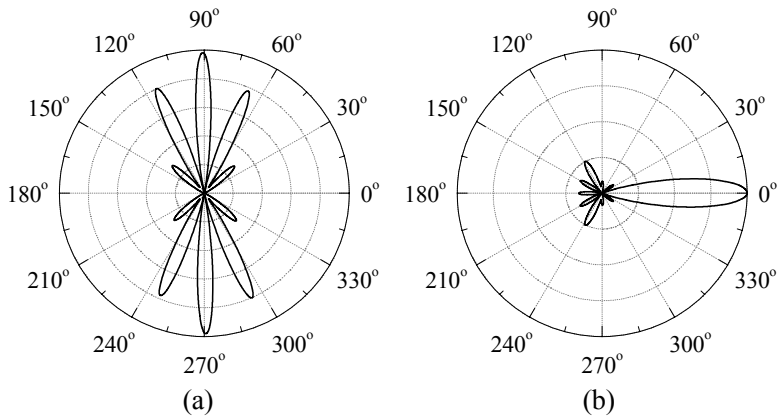

*Figure 9.* Directional far-field emission patterns of the WG modes in (a) an elliptical (Boriskina, 2003) and (b) a notched (Boriskina, 2005) microdisk resonator.

If, as a result of optical or electrical pumping, the optical gain in the resonator active material begins to dominate optical loss, stimulated emission in the microresonator begins to dominate spontaneous emission. To improve the microlaser efficiency, it is desirable to reduce the threshold input energy at which lasing (stimulated emission) begins to occur. The laser threshold depends on the size of a laser microresonator, the lasing mode optical confinement, the gain/absorption balance of the resonator material and the presence of several competing modes within a material spontaneous emission range. Reducing the number of available modes in the spontaneous emission range would therefore both reduce the threshold and improve the laser noise characteristics. This can be achieved by either detuning the wavelengths of all the modes but one against the spontaneous emission peak, or decreasing their Q-factors and thus reducing spontaneous emission into these modes. For example, all the higher-radial-order WG modes in microdisk resonators, with the fields occupying large areas inside the disk, can be suppressed by removing material from the resonator interior (Fig. 8a) by piercing holes (Backes, 1999) or forming microring resonators (Hagness, 1997; Boriskina, 1999, 2002). Enhancement of the working mode Q factor in a microgear (circular microdisk with a periodically corrugated rim) laser has also been achieved (Fujita, 2002), together with the suppression of the Q factor of the other mode of a doublet appearing due to splitting of a double-degenerate first-radial-order WG mode of a circular microdisk (Fujita, 2001; Boriskina, 2004; Fig. 8b). Another way to reduce the threshold in a microcavity laser is to increase the fraction of spontaneous emission into the lasing mode by improving the spectral alignment between the material gain peak and the microcavity mode wavelength (Fujita, 2001; Cao, 2003).

Finally, most practical optoelectronic applications require light sources with a directional light output, e.g., for efficient coupling into the narrow acceptance cone of an optical fibre. In the micropost and micropillar resonators, emission occurs in a single-lobe pattern (Gayral, 1998; Pelton, 2002), which makes their coupling to fibres relatively straightforward. However, due to a rotational symmetry of the structure, the emission patterns of the WG modes in microspheres or circular microdisks are not unidirectional. Instead, they have as many identical beams as twice the azimuthal index of the WG mode. One of the ways to directionally extract the light from such microresonators is to use output couplers, e.g., evanescent-field couplers of various configurations described in Section 5.1. Alternatively, microresonators with non-circularly-symmetric geometrical shapes can be designed to obtain directional emission patterns. Directional light output has been observed from deformed spherical resonators; quadrupolar, elliptical (Noeckel, 1994; Gmachl, 1998; Backes, 1999; Lee, 2002; Boriskina, 2002; Kim, 2004; Fig. 9a), and egg-shaped (Shima, 2001; Boriskina, 2002) microdisks; microdisks with patterned gratings and tabs (Levi, 1992); micropost and disk resonators with spiral cross-sections (Chern, 2003; Kneissl, 2004); and microdisks with notches and openings (Chu, 1994; Backes, 1999; Boriskina, 2006; Fig. 9b).

## 6.  SIMULATION METHODS

All the above-discussed technological advancements in microresonator fabrication and functional applications call for innovative computational methods for solving Maxwell's equations, which govern the electromagnetic fields in the resonator structures. Development of rigorous yet efficient simulation methods and CAD tools is necessary to fully exploit the potential of the new generation of microresonators, substantially reducing the cost and time of the design stage, to create novel devices, and to study new optical phenomena in microcavities.

Unfortunately, Maxwell's equations can be solved analytically for only a few simple canonical resonator structures, such as spheres (Stratton, 1997) and infinitely long cylinders of circular cross-sections (Jones, 1964). For arbitrary-shape microresonators, numerical solution is required, even in the 2-D formulation. Most 2-D methods and algorithms for the simulation of microresonator properties rely on the Effective Index (EI) method to account for the planar microresonator finite thickness (Chin, 1994). The EI method enables reducing the original 3-D problem to a pair of 2-D problems for transverse-electric and transverse-magnetic polarized modes and perform numerical calculations in the plane of the resonator. Here, the effective refractive index of the 2-D structure is taken as the normalized propagation constant of the fundamental guided mode in an equivalent planar waveguide with the same thickness and material as the microresonator.

The finite-difference time-domain (FDTD) method (Chu, 1989) has been a main workhorse in modelling and design of optical microresonators (both in 2-D and 3-D) over the past few decades, due to its simplicity and flexibility (Fujita, 2001; Fong, 2003, 2004; Ryu, 2004; Painter, 2001; Pelton, 2002; Hagness, 1997; Vuckovic, 1999; Rosenblit, 2004). However, the FDTD method suffers from relatively large dispersive error, staircasing errors at resonator boundaries, and reduction of accuracy to the first order at material interfaces. Furthermore, optical microresonators are often placed into complex open space domains, in which case discretization of the problem can lead to errors caused by non-physical backreflections from the edges of the computational window. For large and complex domains, and especially in full-vectorial 3-D simulations, development of faster algorithms becomes imperative to reduce the cost of computations, as the grid size required by using the FDTD method becomes prohibitively expensive for even modern advanced computers. For example, CPU times for the FDTD computations of the characteristics of a 2.25 $\mu$m-diameter semiconductor microring resonator evanescently coupled to a straight semiconductor waveguide reached 35 hours on a 1-GHz, 4-GB RAM platform (Fujii, 2003).

By using fast algorithms based on approximate techniques such as geometrical optics, "billiard theory", paraxial approximation, etc. useful insight into the ray dynamics within optically large microresonators can be obtained (Chowdhury, 1992; Noeckel, 1994, 2000; Jiang, 1999; Poon, 2001; Guo 2003; Huang 2001; Fong, 2004). However, the accuracy of such methods is clearly not adequate for studying the modal spectra of microresonators whose dimensions are of the order of an optical wavelength. Among other popular first-order numerical tools for the microresonator design are: algorithms based on the Coupled Mode Theory (Yariv, 1973; Little, 1997; Manolatou, 1999; Chin, 1998, Hammer, 2002), which are able to provide fast initial designs of waveguide-coupled circular and square microdisk resonators; and the Spectral Index method (Greedy, 2000), which has been successfully applied to study the WG mode characteristics of high-index-contrast circular microdisk resonators mounted on substrates (Fig. 10).

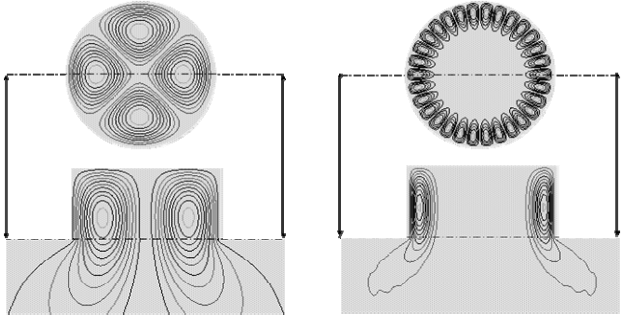

*Figure 10.* Field distributions for two WG modes in a microdisk resonator on a substrate predicted by the Spectral Index method (Greedy, 2000).

Accurate yet computationally efficient techniques that have been already demonstrated to yield solutions to a variety of design problems in optics and photonics are based on the formulation of the problem in terms of either volume or surface integral equations (IEs). A clear advantage of such formulations is that the problem size is reduced by moving it from the open infinite domain to a finite one (a volume or a surface of the microresonator), which reduces the required computational effort. Furthermore, the use of artificial absorbing conditions is completely avoided, together with the danger of unwanted non-physical back-reflections, and microresonators located in the layered media can be modelled without loss of accuracy (Boriskina, 1999, 2002, 2003; Chremmos, 2004). The IE formulation of Maxwell's equations, in its continuous form, is exact provided that the full equivalence takes place. By using various Galerkin-type discretization schemes with low- and high-order basis functions (Harrington, 1968), numerical algorithms with various convergence rates (and thus various computational costs to reach a desired accuracy) can be constructed, with higher-order schemes providing the most computationally efficient solutions (Atkinson, 1997; Nosich, 1999). Due to the flexibility of the method, it has been successfully applied to study modal characteristics of symmetrical and asymmetrical microresonator structures in 2-D, such as circular, elliptical, hexagonal, stadium, and notched microdisk resonators (Wiersig, 2003; Lee, 2004; Boriskina, 1999-2005; Kottmann, 2000, Chremmos, 2004). In 3-D, IE techniques have been developed mainly to analyse dielectric resonators of electrically small size typical for microwave applications (Kucharski, 2000; Glisson, 1983; Umashankar, 1986; Liu, 2004).

It should be noted, however, that the Q factors of open microcavities do not characterise directly the threshold gain values of the corresponding semiconductor lasers. To overcome this difficulty a new lasing eigenvalue problem (LEP) was introduced recently (Smotrova, 2004). The LEP enables one to quantify accurately the lasing frequencies, thresholds, and near- and far-field patterns separately for various WG modes in semiconductor laser resonators. However, the threshold of a lasing mode depends on other factors, including presence of several competing modes within a material spontaneous emission range and detuning of the lasing wavelength against the spontaneous emission peak. To account for these effects, a spontaneous emission coupling factor (β factor) of a given mode can be calculated, which is defined as the ratio of the spontaneous emission rate into that mode and the spontaneous emission rate into all modes (Bjork, 1991). Various approaches to calculate the β factor have been described in the literature, including classical and quantum mechanical methods as well as FDTD (Bjork, 1991; Chin, 1994; Vuckovic, 1999; Xu, 2000).

## 7. CONCLUSIONS

Recent advances in the nanofabrication technology offer the possibility of manufacturing novel optical microresonator devices with unprecedented control in a variety of natural and artificial material systems. High-Q micro- and nano-scale resonators have already proven to be excellent candidates for optical signal processing in low-cost high-density photonic integrated circuits, significant enhancement of detection sensitivity in biosensors, and compact and efficient laser sources with enhanced functionality. Continued development of new materials with specially designed properties such as quantum wires and dots as well as 3-D photonic crystal structures is expected to yield more ideal high-performance optical resonators. These novel resonator designs will fuel future growth of broadband all-optical networks and will impact basic research in quantum physics and quantum information science. However, emerging resonator-based devices and applications as well as growing industrial competition impose strict requirements on the accuracy and performance of the existing resonator design and simulation tools calling for the development of advanced methods and algorithms.

## ACKNOWLEDGEMENT

The authors would like to express their gratitude to the UK Engineering and Physical Sciences Research Council (EPSRC) and the Royal Society for financial support.